\documentclass[twocolumn,showpacs,preprintnumbers,amsmath,amssymb]{revtex4}
\usepackage{epsfig}
\usepackage{graphicx}
\usepackage{dcolumn}

\newcommand{\beq}{\begin{equation}} \newcommand{\eeq}{\end{equation}}
\newcommand{\beqa}{\begin{eqnarray}}
\newcommand{\eeqa}{\end{eqnarray}} \newcommand{\lam}{\lambda}
 
\newcommand{\ga}{\gamma} 
 
 \newcommand{\si}{\sigma}
 \newcommand{\om}{\omega}

\def\oc#1{{ Opt.\ Commun.} {\bf#1}} 
 \def\jpa#1{{ J.\ Phys.\ A} {\bf#1}}
\def\pra#1{{ Phys.\ Rev. A\/} {\bf#1}} \def\prb#1{{ Phys.\ Rev. B\/} {\bf#1}}
 \def\prl#1{{ Phys.\ Rev.\
Lett.} {\bf#1}}

\begin{document}



\title{Quantum Open System Theory: Bipartite Aspects}

\author{T.\ Yu}
\email{ting@pas.rochester.edu}

\author{J.\ H.\ Eberly}
\email{eberly@pas.rochester.edu} 

\affiliation{ Rochester Theory Center for Optical Science and
Engineering, and  Department of Physics and Astronomy, University
of Rochester, New York 14627, USA }


\date{August 30, 2006}

\begin{abstract}

We demonstrate in straightforward calculations that even under
ideally weak noise the relaxation of bipartite open quantum systems
contains elements not previously encountered in quantum noise
physics. While additivity of decay rates is known to be generic for
decoherence of a single system, we demonstrate that it breaks down
for bipartite coherence of even the simplest composite systems.
\end{abstract}


\maketitle


In this note we will establish results that contradict the 
long-standing belief that additivity of coherence decay rates is a 
natural consequence of weak noises. This belief means that the 
relaxation rate of any system exposed to a collection of weak noises 
is the sum of the relaxation rates associated with the noises 
separately. Although implied in many textbook discussions, an actual 
proof of additivity may be difficult to locate.  We supply here a proof of additivity for
a single qubit  coupled to two independent weak noises (here amplitude noise and
phase noise),  but our main message is the 
demonstration of violations of  additivity in the case of entanglement decay 
when two or more qubits are involved.  That is, we will show that a quantum system with the 
most elementary composite structure (e.g., simply made of two 
distinct parts) need not and generally does not exhibit 
relaxation-rate additivity even though the separate parts do. This 
result is purely quantum mechanical and extends our understanding of 
the power of quantum coherence in an unexpected direction.

We now present an additivity proof that is quantum mechanical in
order to eliminate from concern the possibility that quantum systems
are intrinsically different from classical ones in their response to
weak noise. We will consider ideal noise sources where ``ideal" means
that the noise is sufficiently weak and random that the noise-system
interaction is both reliably linear and without significant back
action on the noise source. Each noise source can then be treated as
a reservoir made of an infinite collection of random and very
broadband harmonic oscillators at zero temperature.

Of course the relaxing system need not be linear, so we choose the
simplest nonlinear system, a qubit (two-level atom, spin one-half,
etc.), for our example. The total Hamiltonian for a qubit coupled to
two noise sources (two ``environments") can be written as follows:
\beq
\label{hamiltonian}
H_{\rm tot} = H_{\rm sys} + H_{\rm int} + H_{\rm env},
\eeq
where (with $\hbar =1$)
$$H_{\rm sys} = \frac{1}{2}\omega_0 \si_z \quad {\rm and} \quad
H_{\rm env } = \sum_{\lambda}\omega_{\lambda} a_{\lambda}^\dag
a_{\lambda} + \sum_{\mu}\nu_{\mu}b_{\mu}^\dag b_{\mu}$$
are the Hamiltonians of the qubit system and two local environments.
As an example of the types of relaxation that will be relevant, we
suppose the two environments couple in the one case longitudinally
and in the other case transversely to the qubit. Thus we have for the
interaction of the qubit with its two different noise sources:
\beqa
H_{\rm int}
&=& k_1 \sum_{\lambda} (f^*_{\lambda} \sigma_- a_{\lam}^\dagger +
f_{\lambda} \sigma_+ a_{\lambda}) \nonumber \\
&&+ k_2 \sum_{\mu} \sigma_z (g^*_{\mu}b_{\mu}^\dagger + g_{\mu} b_{\mu}).
\eeqa
We naturally assume that the two noise sources are not
cross-correlated, and in the ideal case under consideration they can
be treated in the familiar Born-Markov limit. Thus we write the
longitudinal and transverse self-correlation functions in the form
$\alpha_{1}(t,s) = \Gamma_1 \delta(t-s)$ and
$\alpha_{2}(t,s) = \Gamma_2\delta(t-s)$,
respectively. The calculation of the time dependence of qubit
coherence follows usual rules \cite{GSA, MOS-MSZ} and we find, for
longitudinal noise alone $(k_1=1, k_2=0)$,
$$\rho_{12}(t) = e^{-i\omega_0t} e^{-\frac{1}{2}\Gamma_1 t}\rho_{12}(0),$$
while for transverse noise alone ($k_1=0,k_2=1$), we have
$$\rho_{12}(t) = e^{-i\omega_0t} e^{-\Gamma_2t}\rho_{12}(0).$$

Now we switch on both longitudinal and transverse noise at the same
time ($k_1 = k_2 = 1$).  The master equation for the qubit system
after tracing over two noise variables is simply given by (in the
interaction picture)
\beqa
\frac{d}{dt}\rho
& = & \frac{\Gamma_1}{2}(2\sigma_- \rho\sigma_+ -\sigma_+
\sigma_-\rho - \rho\sigma_+\sigma_- )\nonumber\\
&&+ \frac{\Gamma_2}{2}(\sigma_z \rho \sigma_z - \rho).
\eeqa

The explicit solution of the above equation gives, for the qubit coherence,
\beq
\rho_{12}(t) = e^{-i\omega_0t} e^{-(\frac{1}{2}\Gamma_1 + \Gamma_2)t}
\rho_{12}(0).
\eeq
This is all that is needed for a proof that the total internal
decoherence rate of a qubit under ideal longitudinal and transverse
noises applied at the same time is given by the sum of the separate
rates: $\frac{1}{2}\Gamma_1 + \Gamma_2$. Finally, note that the
linearity of the ideal noise interactions makes it obvious that any
number of sources of longitudinal noise (any number of distinct
$f_\lam^{(n)} a_\lam^{(n)}\sigma_+$ terms) will additively contribute
to a total $\Gamma_1$, and similarly all
$g_\mu^{(m)}b_\mu^{(m)}\sigma_z$ transverse noise sources will
contribute to $\Gamma_2$.

To the present time treatments of open quantum system theory 
\cite{open2, open1, open3} are based on this scenario in which a 
``small" system has a weak interaction with one or more reservoirs, 
and this is the cause of its relaxation (its loss of self-coherence). 
Now we extend the discussion very slightly, and consider in detail 
the simplest quantum system made of two parts, a pair of qubits. 
Remarkably, this simple step takes us onto new ground within the 
theory of quantum open systems. We will show that the internal 
coherence of the two-qubit system exhibits a non-additive response. 
We believe this is the first demonstration of the effect.

In order to ensure focus on the main point, in the following we will 
not permit the two qubits to interact or communicate with each other, 
and will allow them to be influenced only by noise sources that also 
have no contact with each other. The only connection between the 
parts of the two-qubit system will be pure information. Thus the 
Hamitonian for the two qubit case is simply the addition of the 
Hamiltonians (\ref{hamiltonian}) for the two qubits, respectively. 
Two-party aspects of quantum information such as mixed states and 
entanglement are not present in any single system or in any pair of 
classical systems and can lead to new open system effects. Although 
more general results can be obtained using our methods, we will 
concentrate on mutual entanglement as the most useful measure of 
bipartite coherence for our demonstration. To determine entanglement 
quantitatively we will use concurrence \cite{Wootters}.

Solutions of the appropriate (Born-Markov) equations for noisy
evolution of two-qubit density matrices can be obtained via several
routes \cite{BMLexamples}, and we find the Kraus operator form
\cite{Kraus} most convenient. Given a state $\rho$ (pure or mixed),
its evolution can be written compactly as
\beq \label{Kraus}
\rho(t) = \sum_\mu K_\mu(t)\rho(0) K_\mu^\dag(t),
\eeq
where the so-called Kraus operators $K_\mu$ satisfy $\sum_\mu
K_\mu^\dag K_\mu = 1$ for all $t$.

In order to demonstrate the breakdown of additivity for ideally weak 
noises, it suffices to find a two-party state that experiences 
continuous exponential decay under each noise, but fails to do so 
when  two or more noises are active at the same time. Actually, we 
can identify an entire class of such states. What is more, the class 
is widely known to be relevant in a variety of physical situations 
including pure Bell states \cite{bell} and the Werner mixed state 
\cite{Werner} as special cases.

This class of bipartite states is represented by the following 
two-qubit density matrix, where we use conventional ordering of rows 
and columns related to eigenstates of $\sigma^A_z$ and $\sigma^B_z$ 
in the sequence $[++,~+-, ~-+, ~--]$: \beq \label{e.oldrho}
\rho^{AB} =
\begin{pmatrix}
a & 0 & 0 & 0 \\
0 &  b &  z & 0\\
0 &  z^* &  c & 0\\
0 & 0 &  0 & d
\end{pmatrix}.
\eeq
Obviously $a+b+c+d = 1$. We easily find the concurrence of this state to be
given by
\beq \label{concurrence}
C^{AB} = 2 \max\{0, |z| - \sqrt{ad} \}.
\eeq

For even greater simplicity within this set of density matrices we
will first focus on a smaller sub-category with a single positive
parameter $\lambda$:
\beq \label{e.alteredrho}
\rho_\lam^{AB} = \frac{1}{9}
\begin{pmatrix}
1 & 0 & 0 & 0 \\
0 &  4 &  \lambda & 0\\
0 &  \lambda &  4 & 0\\
0 & 0 &  0 & 0
\end{pmatrix}.
\eeq
Initially for $\rho^{AB}_\lam$ we have $C_\lam(0) = 2\lam/9$.

To begin our time-dependent calculations, we consider pure transverse
(phase) noise, for which we have the following compact Kraus
operators for independently evolving qubits $A$ and $B$:
\begin{eqnarray}
             \label{k1}K_1
&=&\left(\begin{array}{clcr}
\gamma_A && 0\\
0 && 1\\
\end{array}
              \right)\otimes  \left(
\begin{array}{clcr}
\ga_B & 0\\
0 & 1\\
\end{array}
              \right),\\
              K_2&=&\left(\begin{array}{clcr}
\gamma_A && 0\\
0 && 1\\
\end{array}
              \right)\otimes\left(
\begin{array}{clcr}
0 & 0 \\
0 & \om_B\\
\end{array}
              \right),\\
                      K_3&=& \left(
\begin{array}{clcr}
0 & 0\\
0 & \om_A \\
\end{array}
              \right) \otimes \left(
\begin{array}{clcr}
\ga_B & 0\\
0 & 1\\
\end{array}
              \right),\\
                      K_4 &=& \left(
\begin{array}{clcr}
0 & 0\\
0 & \om_A\\
\end{array}
              \right)\otimes \left(
\begin{array}{clcr}
0 &  0 \\
0 & \om_B\\
\end{array}
              \right),\label{k5}
              \end{eqnarray}
where the time-dependent Kraus matrix elements are
$$\gamma_A(t) = \exp{[-\Gamma^A_{2} t/2]} \quad {\rm and} \quad
\om_A(t) = \sqrt{1-\gamma^2_A(t)},$$
and similar expressions for $\gamma_B(t)$  and $\om_B(t)$. We take
$\Gamma^A_{2} = \Gamma^B_{2} \equiv \Gamma_{2}$ for greatest
simplicity. We note that both of the mixed states written above have
the property that they retain their form under these Kraus operators.
For pure dephasing noise the diagonal elements are constant ($a(t) =
1/9, d(t) = 0)$ and the Kraus operators give  $z = \frac{\lam}{9} \to
z(t) = \frac{\lam}{9}\exp[-\Gamma_{2} t]$, and then the phase-noise
concurrence decays asymptotically  exponentially:
\beq \label{phasedecay}
C^{\rm ph.}_\lam(t) = (2\lam/9)\exp[-\Gamma_2 t].
\eeq

\begin{figure}[!t]
\includegraphics[width=6 cm]{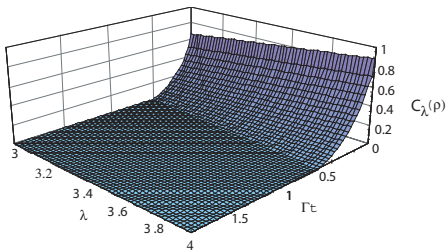}
\caption{{\footnotesize \label{figESD}  The graph shows  $C_\lam$ vs.
$\lambda$ and $t$ under the influence of combined phase and amplitude
noise. The consequence is that (\ref{e.alteredrho})
disentangles completely and abruptly in just a finite time for all
$\lam$ in the range shown.  }}
\end{figure}

For longitudinal (amplitude) noise, the Kraus operators are slightly
different (see \cite{Yu-Eberly04}), but a direct calculation for the
one-parameter example above gives
\beqa
z = \frac{\lambda}{9} \to z(t) &=& \frac{\lambda}{9}\exp[-\Gamma_1 t],  \\
a =\frac{1}{9}\to a(t) & = & \frac{1}{9}\exp[-2\Gamma_1 t], \\
d=0 \to  d(t) &=&  \frac{1}{9}\om_1^4 + \frac{8}{9}\om_1^2,
\eeqa
where $\om_1 = \sqrt{1-\exp[-\Gamma_1 t]}$ and  $\Gamma_1$ is the
longitudinal decay rate for amplitude noise. From these one easily
shows that in the range $4 \ge \lambda \ge 3$ the amplitude-noise
concurrence is given by
\beq \label{ampldecay}
C^{\rm am.}_\lam(t) = \frac{2}{9} \left[\lam - \sqrt{\om_1^4 +
8\om_1^2}\right] \exp[{-\Gamma_1 t}].
\eeq
Therefore, our bipartite entanglement under amplitude noise also
decays smoothly and asymptotically exponentially to zero. With these
two exercises in hand we conclude that for our mixed two-party system
the entanglement decays asymptotically smoothly to zero in the
presence of either weak amplitude noise or weak phase noise.

Now we consider the issue of additivity and allow the weak noises to 
be applied together. All two-party density matrix elements decay at 
the sum of their respective phase and amplitude rates, as the 
additivity theorem ensures. However, for the entanglement 
measure of coherence the consequence is strikingly different. 
Because the off-diagonal element $z=\lam/9$ decays at the sum of the 
separate phase- and amplitude-noise rates, the two-noise concurrence 
takes the form:
\beq \label{two-noise concur}
C^{\rm ph.+am.}_\lam(t) = 2\max\{0, \lam e^{-\Gamma_2t} -
\sqrt{\om_1^4 + 8\om_1^2}\},
\eeq
which has lost any trace of relaxation additivity, particularly the
property of an asymptotically smooth approach of entanglement to
zero. As we show in Fig. \ref{figESD}, over a continuous range of
physical $\lam$ values, $C^{\rm ph.+am.}_\lam(t)$ actually goes abruptly
to zero in a finite time, and remains zero thereafter. This is the
effect that has been called ``entanglement sudden death"
\cite{Yu-Eberly06}, and it arises here more or less from nowhere,
since there is no local effect, under the action of either of the
weak noises, indicating that it should be expected. The present
result shows that ESD is one consequence not previously noted,
indicating necessary departures from standard elements of open-system
theory in multi-party relaxation, even for ideally weak noise
influences.

\begin{figure}[!b] 
\includegraphics[width=8 cm]{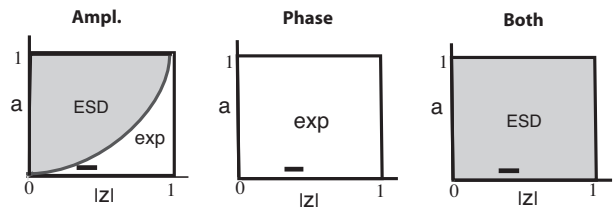}
\caption{{\footnotesize \label{fig3}  This diagram shows the dramatic
effect of the combination of two noises, amplitude and phase noise in
this case, on all initial states (\ref{e.oldrho}) with $d=0$.  (i)
Amplitude noise can lead to entanglement sudden death (dark zone),
but for a large parameter range (white zone, $a \le |z|^2$) the
entanglement only decays exponentially. (ii) Under phase noise, the
initial entanglement always decays exponentially (white zone). (iii)
However, when the noises are combined, {\em all initally entangled
states} suffer sudden death (dark zone).  In each part the solid-line
segment ($a=1/9, 1/3\leq |z| \leq 4/9$) shows the parameter range
associated with the particularly simple concurrence $C_\lam$  that we
discussed in detail. }}
\end{figure}

It is important to emphasize that our special one-parameter example
is not a singular case. The simplest verifications of this can be
made by just retaining $d=0$ within the more general matrix class
(\ref{e.oldrho}). The outcome of fairly straightforward calculations
for the entire class is illustrated by the diagrams in Fig.
\ref{fig3}. Part (i) shows that under pure amplitude noise either ESD
or pure exponential decay may occur, with the boundary between them
given just by $a = |z|^2$. Part (ii) indicates that for the entire
range of $a$ and $|z|$ under pure phase noise the decay is purely
exponential.  However, (iii) shows that  under the combination of
phase and amplitude noise every initial state (\ref{e.oldrho}) will
disentangle abruptly. This directly shows that when the parameters
lie in the zone $a \le |z|^2$,  non-additivity occurs for
entanglement decay rates.

We end our examination with a general observation. The  calculations
displayed here  reach only a small corner of a new domain of noise
physics. Wider questions can also be answered. What if one  applies
both phase noise and amplitude noise to sub-system $A$ alone, leaving
$B$ totally noise-free? One finds that this is enough to impact the
bipartite $AB$ entanglement just as strongly as before. What if one
applies only phase noise to sub-system $A$ and only amplitude noise
to $B$? In that case both $A$ and $B$ have to relax normally, but
their mutual entanglement does not. These results can be verified by
straightforward calculations. The fact that the same conclusion
applies no matter where one looks in this domain demonstrates that 
information about an
open bipartite quantum system will become degraded with time as an 
indivisible quantum unit,
no matter how its parts are engaged by weak noises, and the 
degradation is not predicted by the familiar smoothly decaying
behavior familiar from the quantum theory of single open systems.

To summarize, in this note we introduced a commonly occurring 
category of two-system mixed states, shown in (\ref{e.oldrho}). By 
following their time dependent behavior under the influence of 
ideally weak noises we demonstrated the presence of elements of open 
quantum system theory not previously encountered. These become 
interesting whenever a small system has different quantum parts that 
can be entangled. Exactly this situation will arise, for example, in 
a quantum computer, where it is most desirable that two qubits retain 
a non-zero degree of mutual cross-entanglement. It must be emphasized 
that none of our key $AB$ results come from interaction or 
communication between the $A$ and $B$ parts of the two-party system, 
or between their separate reservoirs.

This is perhaps the most striking aspect of the properties described: 
they are properties of joint-system information rather than 
joint-system interaction. To the extent that joint system information 
is a resource of substantial value in one or another practical 
application of qubit networks, this aspect of time-dependent 
entanglement will be important. At the same time it illuminates 
further the difficult fundamental challenge to understand the nature 
of coherence in multi-partite mixed states, particularly in its 
time-dependent behavior, which has recently come under examination in 
both continuous spaces \cite{Diosi, Halliwell-etal,Ban} and discrete 
spaces (qubit pairs \cite{Yu-Eberly02, Yu-Eberly03,Malinovsky, 
Kamta-Starace, arXiv1}, finite spin chains and elementary lattices 
\cite{Pratt-Eberly01, Pratt04a, Carvalho-etal, privman2}), and 
decoherence dynamics in adiabatic entanglement \cite{Sun01}, as well 
as in situations without relaxation \cite{Yonac-etal06} and in 
connection with direct entanglement observation \cite{Santos-etal06}. 
These have all contributed to increased awareness of this domain.

Finally, it should be emphasized that although entanglement  measured by concurrence 
is not an observable represented by an Hermitian operator, nevertheless  it is still possible to 
express the concurrence (\ref{concurrence}) in terms of the expectation values of certain ordinary
physical observables  \cite{notes}.  Moreover,  the recent proposals to directly 
measure the dissipative entanglement evolution have opened up a possibility 
of experimentally demonstrating the onset of the non-additivity when nonlocal 
coherence decay is concerned \cite{Santos-etal06,nature}.

We acknowledge financial support from NSF Grants
PHY-0456952 and PHY-0601804, and ARO Grant W911NF-05-1-0543.


\begin{thebibliography}{99}
\bibitem{bell} The matrix (\ref{e.oldrho}) can give rise to all four
Bell states by local operations. Notably, a more general density
matrix with extra non-zero elements $\rho_{14}$ and $\rho_{41}$ will
also allow analytical solutions. In this paper we have restricted our
discussion to the form given by (\ref{e.oldrho}).
\end{thebibliography}
\end{document}